\documentclass[9pt,twocolumn]{extarticle}
\usepackage[utf8]{inputenc}
\usepackage[T1]{fontenc}
\usepackage{amsmath,amssymb,amsfonts}
\usepackage{hyperref}
\usepackage{fancyhdr}
\usepackage{authblk}
\usepackage{enumitem}
\usepackage{booktabs}
\usepackage{siunitx}
\usepackage[nolist]{acronym}
\usepackage[format=plain,font=it]{caption}
\usepackage{xcolor}
\usepackage{cite}

\hypersetup{
	colorlinks,
	linkcolor={red!50!black},
	citecolor={blue!50!black},
	urlcolor={blue!80!black}
}

\AtBeginDocument{%
  }

\usepackage[final]{changes}
\usepackage{geometry}
\begin{acronym}[ABCD]
	\acro{APF}{Annual Performance Factor}
	\acro{COP}{Coefficient of Performance}
	\acro{EU}{European Union}
	\acro{EV}{Electric Vehicle}
	\acro{HP}{Heat Pump}
	\acro{HVAC}{heating, ventilation and air-conditioning}
	\acro{IoT}{Internet of Things}
	\acro{JSD}{Jensen-Shannon Divergence}
	\acro{PV}{Photovoltaic}
	\acro{MILP}{Mixed-Integer Linear Program}
	\acro{SPF}{Seasonal Performance Factor}
\end{acronym}

\title{Modeling of Annual and Daily Electricity Demand of Retrofitted Heat Pumps based on Gas Smart Meter Data}

\fancypagestyle{firstpageheader}{%
	\fancyhf[L]{\textit{Accepted version of the paper submitted to the 10th ACM International Conference on Systems for Energy-Efficient Buildings, Cities, and Transportation, November 15--16, 2023, Istanbul, T{\"u}rkiye (BuildSys '23)}}
	\fancyhf[R]{}
	\fancyfoot[L]{\textit{DOI of the published version: \href{https://dx.doi.org/10.1145/3600100.3623745}{10.1145/3600100.3623745}\\\copyright~2023 Copyright held by the authors. Publication rights licensed to ACM. ACM ISBN 979-8-4007-0230-3/23/11..\$15.00}}
	\fancyfoot[C]{}
}

\author[1]{Daniel R. Bayer\footnote{Email address: daniel.bayer@uni-wuerzburg.de}~}
\author[1]{Marco Pruckner\footnote{Email address: marco.pruckner@uni-wuerzburg.de}}
\affil[1]{University of W{\"u}rzburg, Am Hubland, W{\"u}rzburg, Germany}
\date{}

\begin{document}

\maketitle
\thispagestyle{firstpageheader}

\begin{abstract}
  Currently, gas furnaces are common heating systems in Europe.
  Due to the efforts for decarbonizing the complete energy sector, heat pumps should continuously replace existing gas furnaces.
  At the same time, the electrification of the heating sector represents a significant challenge for the power grids and their operators.
  Thus, new approaches are required to estimate the additional electricity demand to operate heat pumps.
  The electricity required by a heat pump to produce a given amount of heat depends on the \ac{SPF}, which is hard to model in theory due to many influencing factors and hard to measure in reality as the heat produced by a heat pump is usually not measured.
  % generating the same amount of heat as gas furnaces did.
  Therefore, we show in this paper that collected 
  % heat pump electricity and gas consumption
  smart meter data forms an excellent data basis on building level for modeling heat demand and the \ac{SPF}.
  %In this paper, we exploit smart meter data for ..
  %Nowadays, distribution network operators install more and more smart metering systems collecting the electricity demand of heat pumps or gas demand of gas furnaces with a high time resolution.
  %Up to now, a lot of research has been done on analyzing either heat pump electricity or gas consumption data, without pairing these two data sources.
  %
  %If an individual gas furnace is assumed to be replaced by a heat pump, the question arises how a measured gas profile can be converted to a heat pump profile and how the annual gas consumption can be translated to a heat pump electricity consumption.
  %
  We present a novel methodology to estimate the mean \ac{SPF} based on an unpaired dataset of heat pump electricity and gas consumption data taken from buildings within the same city by comparing the distributions using the \ac{JSD}.
  Based on a real-world dataset, we evaluate this novel method by predicting the electricity demand required if all gas furnaces in a city were replaced by heat pumps and briefly highlight possible use cases.
\end{abstract}

\acresetall % reset all aconyms

\section{Introduction}
	In the next years, the number of installed heat pumps in the residential sector is expected to increase strongly, replacing existing gas furnaces, especially in Europe, in order to decarbonize the building sector \cite{2022_HeatPumpMarketComment}.
	The \ac{EU} pursues this goal politically with vigor \cite{2023_EU_Press_Release}.
	Germany, for example, will reach net climate neutrality by 2045, where heat pumps will be the main driver in the decarbonization of the residential heat sector \cite{2022_DE_ClimateActionInFigures}.
	These new heat pumps will substantially increase the electricity demand, especially at the distribution grid level.
	Thus, utility companies need to predict the impact of these retrofitted heat pumps on their distribution grids.
	Based on this prediction, possible grid bottlenecks and overloads can be identified at an early stage.
	As a result, grid expansion can be carried out early and in a targeted manner.
	
	To predict the retrofitted heat pump impact on the grid, we need information about the heat demand of the existing buildings, which we calculate based on increasingly available smart meter data.
	Gas consumption data can be converted on the building level to its heat demand, given the furnace efficiency and the gas heating value - both daily or annually.
	%Due to the ongoing development of the \ac{IoT}, the collection of electricity and gas consumption metering data becomes easier and is widely utilized.
	%Smart meter data becomes more and more available from both, electricity and gas meters.
	%For example, in the United Kingdom, \num{28.1} million smart meters were installed until end of 2022, which is about 50\% of all households \cite{2022_UKSmartMeterReport_Q4}.
	%For example, in the United Kingdom, smart meters were installed in about 50\% of all households until end of 2022 \cite{2022_UKSmartMeterReport_Q4}.
	%
	%The data of those smart meters can be used for multiple tasks, like load profiling \cite{2018_Kahn_LoadProfilingUsingSmartMeterData}, load forecasting \cite{2020_STLoadForecastingUsingSmartMeterData} or building a digital twin \cite{2023_Bayer_DT_LocalEnergySystem}.
	%This data gets very valuable in view of the described changes that will happen in the energy sector.
	%This data gets very valuable when predicting the retrofitted heat pumps electricity demand.
	%In the context of gas furnace replacement the question arises how we can estimate the electricity consumption of a heat pump that should replace an existing gas furnace.
	%Converting an annual or daily gas consumption to the building heat demand is trivial given the furnace efficiency and the gas heating value.
	To compute the electrical energy consumption of a heat pump that should generate the same amount of heat, we need to know the (heating) \ac{SPF} that is highly dependent on the climatic conditions where the heat pump is installed \cite{2015_Nouvel_SPF_in_Europe}.
	Existing approaches like \cite{2023_DecarbonizationOfResidentialHeatingBasedOnSMD} use average values for \ac{SPF} (or \ac{COP}).
	However, in a current grid state, we will find meter data of new, already existing heat pump installations besides the meter data of older gas furnaces.
	This paper addresses the issue of estimating the \ac{SPF} based on an unpaired smart meter data set of existing heat pumps and gas furnaces.
	\deleted{
		For validation, we pair this data with the building volume, as this parameter can easily be inferred from city-wide 3D building models \cite{2011_3DBuildingModel_Bavaria}.
	}
	%The building volume is the parameter with the highest influence on individual building thermal energy usage (in the residential sector) \cite{2013_Ko_ReviewDesingResearchResidentialEnergyUsage}.
	%Our basic idea is to use the recorded smart meter data of gas meters for estimating the electricity demand of heat pump if the existing gas furnace is replaced by a heat pump.
	%Our basic idea is to estimate, based on recorded gas and electricity smart meter data, what the energy consumption of a heat pump would be if an existing gas heating system were replaced by a heat pump.
	%Our basic idea is to use the data of existing consumers, i.e., heat pumps, to model similar heat pumps that will be installed somewhere in the future.
	%
	In detail, we answer the following research questions in this paper:
	\begin{enumerate}
		\item How can we predict the mean \ac{SPF} over all existing heat pump installations based on an unpaired heat pump electricity and gas consumption dataset?
		\item How much electricity would be consumed by heat pumps that replace existing gas furnaces based on the previously estimated \ac{SPF}?
	\end{enumerate}
	%To answer these questions, 
	%Additionally, we present an analysis of the daily gas consumption compared to the daily electricity consumption of the heat pumps already installed.
	In order to address these research questions, we present a novel approach using the \ac{JSD} to compare the heat pump electricity and gas furnace consumption distributions.
	%Finally, we discuss use cases and benefits of this novel approach.
	%Our analysis is based on data we got from the Stadtwerk Ha{\ss}furt GmbH, a local distribution network operator located in Ha{\ss}furt, Germany.

	The subsequent parts of this paper have the following structure.
	In \autoref{sec:related_work}, we present related work dealing with similar research questions.
	We follow with details on our methodology in \autoref{sec:methodology}.
	The results are presented in \autoref{sec:results}.
	Subsequently, we discuss these results and highlight the benefits of this novel approach.
	Finally, we conclude the paper in \autoref{sec:conclusion}.

\section{Related Work}
\label{sec:related_work}

	% data usage of gas or heat pump data
	Real measured gas consumption data is used in \cite{2011_Kolter_PredictingAndVisualizingBuildingEnergyUsage} to predict and visualize the heat demand on the building level.
	The author of \cite{2013_Ko_ReviewDesingResearchResidentialEnergyUsage} presents a review of modeling and statistical analysis of measured heat and energy usage in the urban sector.
	%The authors highlight that most statistical analyses of residential heat demand identify building size as most influencing parameter for annual heat demand.
	Assuming the presence of hourly heat demand data for an existing building in conjunction with weather data, we can use the model presented in \cite{2022_Lumbreras_HeatDemandModel_SMD} for an hourly long-term heat demand prediction.

	% general models for SPF or COP
	Several models exist that estimate either the \ac{COP} \cite{2021_Shin_ML_HP_COP_Model} or the \ac{SPF} \cite{2015_Nouvel_SPF_in_Europe} based on theoretical computations or laboratory measures.
	In the latter, for example, the authors present a model for estimating the \ac{SPF} of a highly simplified sample building placed in different regions of Europe. %, comparing air- and ground-source heat pumps.
	% SPF models or studies based on real data
	An analysis of the electricity demand and \ac{SPF} of real-world heat pump installations is presented in \cite{2014_Miara_HP_SPF_RealMeasurement} and \cite{2012_Huchtemann_FieldTestHPRetrofit}, presenting notable differences from the theoretical computed values.
	A review of field studies of heat pump efficiencies is presented by \cite{2020_Carroll_HP_FieldStudiesReview}.
	
	% these models used for replacing gas furnace by heat pumps
	Only a few recent works use real gas consumption data, assuming these gas furnaces should be replaced by heat pumps.
	For example, the authors of \cite{2023_DecarbonizationOfResidentialHeatingBasedOnSMD} use gas consumption data of a city to compute the residential energy demand on a household level using an average \ac{COP}.
	They focus on selecting regions of the city where heat pumps should replace all existing gas furnaces to save maintenance costs at the gas grid level.
	%The used average \ac{COP} has not been validated.
	In \cite{2023_Wamburu_HP_replacment_with_data}, the authors present an analysis of gas consumption in a residential area based on real data. Additionally, they propose an optimization framework for reducing carbon dioxide emissions.
	% what is missing
	Nevertheless, existing gas consumption data has not yet been combined with heat pump demand data for estimating the impact of heat pump retrofits in future energy system states.

% TODO: TO which definition of SPF (SPF 2 presumably) do we stick to? Different definitions can be found in \cite{2013_Gleeson_HP_EfficiencyFieldTrialsInEurope}

\section{Methodology}
\label{sec:methodology}

	Let $G$ be the set of existing gas furnaces that should be replaced by heat pumps and $H$ be the set of already existing heat pumps.
	Based on the existing gas consumption data of year $j$ we can compute the thermal energy $Q_j(g)$ produced by furnace $g\in G$ similar to \cite{2023_DecarbonizationOfResidentialHeatingBasedOnSMD} using:
	\begin{equation}
		\label{eq:Q_gj}
		Q_j(g) = R_j(g) \cdot z \cdot \nu \cdot \lambda_g
	\end{equation}
	where $R_j(g)$ is the volume of gas consumed in year $j$, $z$ is the pressure factor, $\nu$ is the heating value of the gas and $\lambda_g$ is the efficiency of furnace $g$.
	
	It is possible that a building's insulation level will be improved if a gas furnace is replaced by a heat pump.
	Therefore, we introduce the parameter $\gamma(g)$, which denotes the heat demand reduction that goes along with the replacement of gas furnace $g$.
	The heat demand $Q^\Delta_j(g)$ after the retrofit is thus given by
	\begin{equation}
		\label{eq:Q_new_gj}
		Q^\Delta_j(g) = Q_j(g) \cdot \left(1-\gamma(g)\right)
	\end{equation}

	According to \cite{2021_Hailu_EnergySystemsInBuildings}, we define the \ac{SPF} of a heat pump $h$ in the year $j$ as:
	\begin{equation}
		\label{eq:SPF}
		\mathrm{SPF}_j(h) = \frac{Q_j(h)}{E_j(h)}
	\end{equation}
	where $E_j(h)$ denotes the electrical energy consumed by heat pump $h$ in year $j$ and $Q_j(h)$ denotes the heat (i.e., thermal energy) produced by heat pump $h$ in year $j$.
	We note that the considered heat pumps produce domestic hot water (but no cooling loads) during summer.
	%Thus, the \ac{APF} (see for example \cite{2013_SPF_APF}) equals the \ac{SPF}.

	If a heat pump $h$ replaces furnace $g$, the heat required must be equal, i.e., $Q^\Delta_j(g) = Q_j(h)$.
	Thus, we can compute the electrical energy that \added{heat pump} $h$ would consume by combining Equations (\ref{eq:Q_gj}), (\ref{eq:Q_new_gj}) and (\ref{eq:SPF}):
	\begin{equation}
		\label{eq:E_h.j}
		E_j(h) = Q_j(g) \cdot \underbrace{\frac{1-\gamma(g)}{\mathrm{SPF}_j(h)}}_{=:\;B_j^{-1}}
	\end{equation}

	\added{Let us assume}\deleted{Assuming} that first, the \ac{SPF} divided by one minus the insulation enhancement $\gamma(g)$ ($B_j^{-1}$ in Eq.  (\ref{eq:E_h.j})) is independent of the heat demand $Q_j(h)$,
	\deleted{and second, the distribution of the heat demand over all buildings is independent of the heating system type (except of the scaling), we can compute the expected value of both sides of Eq. (4) as}
	%\added{second, both distributions ${E_j}$ and $f_{Q_j}$ equally distributed}
	\added{and second, the distribution of $B_j$ resp. $B_j^{-1}$ is constant over the building stock for a single year $j$}.
	\added{Then, the following equation holds for the distributions computed over the obtained dataset}
	\begin{equation}
		\label{eq:distributions}
		E_j \sim Q_j \cdot B_j^{-1} \Longrightarrow E_j \cdot B_j \sim Q_j
	\end{equation}
	\added{where the tilde denotes the equality of two distributions.}
	\deleted{where $h(g)$ is the heat pump instance replacing furnace $g$.}
	
	\added{To get the value of $B$ where both distributions of Eq. (\ref{eq:distributions}) overlap best, }
	\deleted{As we can compute both distributions of $E_{H,j}$ and $Q_{G,j}$ out of the obtained data set, }we can compute the \ac{JSD} between $E_{j} \cdot B_j$ and $Q_{j}$ for a high resolved, discrete set of values of $B_j$ and take the $B_j$ minimizing the \ac{JSD}.
	The \ac{JSD} is a symmetric measure for the similarity of two distributions (see \cite{1991JensenShannonDivergence} for theoretical details),
	\begin{equation}
		\mathrm{JSD}(P \parallel Q) = \frac{1}{2} D_{\mathrm{KL}}(P \parallel Q) + \frac{1}{2} D_{\mathrm{KL}}(Q \parallel P)
	\end{equation}
	where  $P$ and $Q$ are two distributions and $D_{\mathrm{KL}}$ denotes the Kullback–Leibler divergence.
	It takes smaller values the more similar the distributions are and reaches 0 if and only if both distributions are equal.
	\deleted{Thus, we find the optimal value of $B$ by minimizing the \ac{JSD}.}
	We \added{explicitly} note that this minimization should be applied for every year $j$ individually, as the \ac{SPF} can differ in different years.
	\deleted{
	Suppose we add an assumption about the reduced heat demand $\gamma$ for possible newer buildings with a heat pump installation.}
	\added{Suppose we add an assumption about the mean reduced heat demand $\gamma(g)$ over all buildings with a heat pump $g$, e.g., based on a heat cadaster.}
	In this case, we can directly estimate the mean \ac{SPF} over all existing heat pumps.
	%
	% validation
	\deleted{
	We validate our approach by analyzing the heat demand per building volume as most statistical analyses of residential heat demand identify building size as most influencing parameter for annual heat demand %\cite{2013_Ko_ReviewDesingResearchResidentialEnergyUsage}
	[8].}
	\added{We validate our approach by comparing our results to field trials for \ac{SPF} computation.}

	\subsection{Dataset}
\begin{figure}[h]
		\centering
		\includegraphics[width=0.96\linewidth]{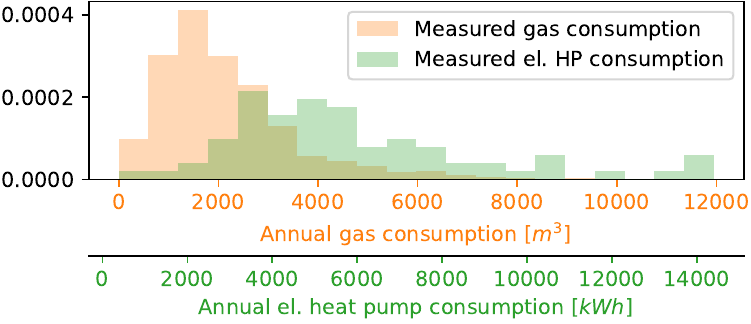}
		\caption{Histogram of total gas (orange) and heat pump electricity (green) consumption from our dataset in 2020.}
		%\Description{A histogram showing the annual gas consumption and the annual electricity consumption of all heat pumps.}
		\label{fig:histograms}
	\end{figure}

	Our analysis is based on about \num{1400} smart meter time series of gas meters with a daily resolution and \num{73} time series of electricity meters exclusively metering a heat pump with an hourly resolution from 2019 until 2021.
	The data was measured in a small town in southern Germany including most of the residential buildings there.
	%We obtain the data from our project partner located in \anon[(city blinded for review)]{Ha{\ss}furt}, Germany.
	For all of the obtained time series, we know the location of the meter and surplus information about the building, like its volume.
	\added{The histogram of the total gas and heat pump electricity consumption in 2020 is depicted in \autoref{fig:histograms}.}
	\deleted{We conducted a survey to identify the share of heating types installed in the town, showing that 76\% of all heat pumps are air-source heat pumps and 24\% are ground-source heat pumps.}
	\added{We conducted a survey among all residential customers of the utility company, which we validated against the building volume of all residential buildings (t(5824) = -0.36, p = 0.72, d = -0.01).
	The response rate was around 17\%.
	Based on the survey results we identified the share of installed heat pump types, showing that 76\% are air-source and 24\% are ground-source heat pumps.}
	
	\paragraph{Dataset validation}
	Our dataset shows a correlation between the mean daily heat pump electricity consumption and the mean daily outside temperature of $r = -0.95$ for 2021 and $r=-0.92$ for 2019 and 2020.
	When comparing the mean daily gas consumption with the mean daily outside temperature, we still get a correlation of $r=-0.92$ for 2020 and 2021 and $r=-0.93$ for 2019.
	These results go along with related publications.
	For example, \cite{2023_Wamburu_HP_replacment_with_data} shows a correlation of $r=-0.90$ between the daily gas consumption and the temperature time series.

	\paragraph{Short analysis of consumption values per building volume}
	\deleted{Figure 1 depicts the correlation of building volume and annual gas consumption for the year 2021 in the upper plot (orange) and the same for the heat pump electricity demand in the lower plot (green).
	We can report a correlation of $r=0.56$ for the annual gas demand in 2021 and a correlation of $r=0.65$ for the heat pump electricity consumption in 2021.}
	\added{We can report a correlation of $r = 0.63$ to $r = 0.65$ between the building volume and both annual values of gas consumption and heat pump electricity consumption for 2019 to 2021.}
	Even though both correlation values show a high deviation, other publications like \cite{2011_Kolter_PredictingAndVisualizingBuildingEnergyUsage} report similar correlation measures ($r=0.61$).
\added{
	Reasons for the high deviation of heat demand have been discussed multiple times \cite{2017_Lim_BuildingStockEnergyPrediction}.
	They are mainly based on unobservable variables like the insulation level, user behavior, non-optimal system configuration or diverse control strategies.
	%Uncertainty may also arise from the diverse control strategies that are applied in practice.
	%For example in \cite{2022_Bayer_Enhancing_RL_for_HVAC}, a controller based on reinforcement learning is able to reduce \ac{HVAC} energy consumption by 6 to 8\%.
	% TODO add possible later
	%For example, a controller based on reinforcement learning can reduce the \ac{HVAC} energy consumption by 6 to 8\%  \cite{2022_Bayer_Enhancing_RL_for_HVAC}.
}
	
	%
	% Daily comparison
	%
	\paragraph{Daily analysis}
	The comparison between the mean daily heat demand produced by all gas furnaces and the mean electricity consumption of all installed heat pumps per day shows a strong correlation of $r=0.99$ in 2019 and $r=0.98$ or $r=0.97$ for 2020 or 2021.

\section{Results}
\label{sec:results}

	Depending on the year of the analysis, different values of $B_j = \frac{\mathrm{SPF}_{j}}{1-\gamma}$ minimize the \ac{JSD}.
	The minimal values are found by computing the \ac{JSD} for a discrete set of values for $B_j$ in a range of $1.5$ to $4.0$ with a step size of $0.1$.
	In 2019, $B = 3.3$ is optimal, $B = 3.5$ in 2020 and $B = 3.0$ in 2021.
	%An overview of the \ac{JSD} values are depicted in \autoref{fig:JSD_plot}.
	\autoref{tab:JSD_temp_values} holds the values of $B$ that achieve a minimal value of the \ac{JSD} for different years in combination with the mean outdoor temperature in the winter (i.e., from 01.01. until 20.02. and from 01.11. until 31.12.).
	The analysis shows an almost linear relationship between the optimal value for $B$ and the mean outdoor temperature in the winter months.
	This seems plausible as our data contains lots of air-source heat pumps that show a reduced efficiency during colder winters \cite{2001_DeSwardt_ASHP_GSHP_Comparison,2013_Vocale_TemperatureImpactOnASHPEfficiency}.
	
	%\begin{figure}[h]
	%	\centering
	%	\includegraphics[width=\linewidth]{figures/JSD_plot_NB039}
	%	\caption{Plot of \ac{JSD} for different values of $B$ from \autoref{eq:E_h.j}.}
	%	\Description{Plot of \ac{JSD} for different values of $B$ from \autoref{eq:E_h.j}.}
	%	\label{fig:JSD_plot}
	%\end{figure}

	\begin{table} % use table* to span across bowth columns
	  \caption{Parameter setting achieving minimal \ac{JSD} values for different years in combination with the mean outdoor temperature in winter}
	  \label{tab:JSD_temp_values}
	  \centering
	  \resizebox{0.9\linewidth}{!}{%
	  \begin{tabular}{ccc}
	  	%\tiny
	    \toprule
	    Year & $B$ with minimal JSD & Mean temperature in winter \\
	    \midrule
		2019 & 3.3 & 3.8\textdegree C\\
		2020 & 3.5 & 4.6\textdegree C\\
		2021 & 3.0 & 2.8\textdegree C\\
	  \bottomrule
	  \end{tabular}}
	\end{table}

	% verification
	% assuming the knowledge of gamma_g
	\added{
	The town council created a heat cadaster to determine the current building insulation in the municipal area.
	Therefore, they randomly selected over 1000 residential buildings and used a questionnaire to evaluate the individual building heat demand and insulation level.
	Based on this cadaster, we can assume that buildings heated with a heat pump have a heat requirement of around 10.5\% lower than buildings currently heated with a gas furnace.
	}
	\deleted{Based on a survey conducted by the town council, we can assume that buildings heated with a heat pump have a heat requirement of around 10.5\% lower than buildings currently heated with a gas furnace.}
	Thus, we can compute the average \ac{SPF} over all buildings to $3.0$ for 2019 and $3.1$ or $2.7$ for 2020 or 2021.
	%These findings fit to real measured \ac{SPF} in the same country.
	%For example, \cite{2012_Huchtemann_FieldTestHPRetrofit} reports a mean measured \acp{SPF} of $2.3$ for air-source and $2.9$ for ground-source heat pumps, installed in 2007.
	%Another heat pump monitoring project in Germany \cite{2014_Miara_HP_SPF_RealMeasurement} reports mean measured \ac{SPF} values of $3.1$ for air-source and $4.0$ for ground-source heat pumps in 2012/2013.
	These findings are in a valid range, as field trials report mean \ac{SPF} of $2.3$ to $3.1$ for air-source and $2.9$ to $4.0$ for ground-source heat pumps in Germany \cite{2012_Huchtemann_FieldTestHPRetrofit,2014_Miara_HP_SPF_RealMeasurement}.
	% fit to real measured \ac{SPF} in the same country.
	%For example, \cite{2012_Huchtemann_FieldTestHPRetrofit} reports a mean measured \acp{SPF} of $2.3$ for air-source and $2.9$ for ground-source heat pumps, installed in 2007.
	%Another heat pump monitoring project in Germany \cite{2014_Miara_HP_SPF_RealMeasurement} reports mean measured \ac{SPF} values of $3.1$ for air-source and $4.0$ for ground-source heat pumps in 2012/2013.
	
\begin{figure}[h]
		\centering
		\includegraphics[width=0.96\linewidth]{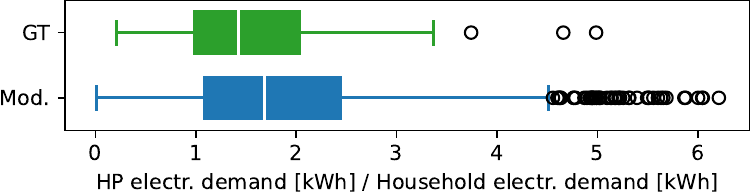}
		\caption{Boxplot of the share of heat pump electricity demand divided by the residential electricity demand (excluding the heat pump) for the year 2021. Green box: Share over the existing heat pump installations. Blue box: share over the future heat pumps (clipped at a ratio of 6.4).}
		%\Description{A boxplot of the share of heat pump demand to household demand for the measured heat pump data and a separate plot for the simulated consumption if gas furnace were replaced by heat pumps. We see very similar boxplots.}
		\label{fig:boxplot_share_HP_to_HH_demand}
	\end{figure}

	\paragraph{Annual demand}
	If all existing gas furnaces in the residential buildings were replaced by heat pumps, and if the insulation level of all gas-heated buildings were equivalent to those with a heat pump, they would have consumed $9.6$ GWh in 2019 and $8.7$ or $10.1$ GWh in 2020 or 2021 in total.
	This corresponds to an increase in electricity consumption over all gas-heated buildings of $149$\% in 2019 and $151$\% or $171$\% in 2020 or 2021.
	\added{
	The ratio of the heat pump to building electricity demand is visualized in \autoref{fig:boxplot_share_HP_to_HH_demand} for 2021.
	The building electricity demand does not contain the heat pump demand.
	Thus, a ratio smaller than one can appear.
	The median of this household-to-heat-pump electricity demand ratio is higher for the future systems replacing gas furnaces (blue) than for the already installed heat pumps (green).
	Additionally, this ratio shows a higher variance for future systems than for existing ones.
	The different variance magnitudes are already present in the initial dataset.
	}
	\deleted{We see, that the ratio based on converted gas consumption data (blue) is in a similar region as the ratio of the already installed heat pumps (green), but shows more variation because we have a much more data points.}
	
\section{Discussion}
\label{sec:discussion}

	Our \added{heat demand analysis} shows a notable variation in the annual heat demand for buildings with the same volume.
	Therefore, the investigation of the annual electricity demand of a potentially retrofitted heat pump should be based on the individual building heat demand, if such data is available (like in \cite{2023_DecarbonizationOfResidentialHeatingBasedOnSMD}), and not only rely on building parameters like its volume.
	The fact that gas consumption and electricity heat pump consumption show a high correlation on a daily level allows us to assume that gas profiles can be easily replaced by heat pump profiles.
	This finding simplifies the modeling of future retrofitted heat pumps.
	Moreover, our estimates for the \ac{SPF} for different years lie in a valid range.
	However, the notable annual deviations argue against using average values for the \ac{SPF}.
\added{
	% Requirements of our method
	Our methodology requires an assumption about the difference in heat demand between the buildings with heat pumps and those with gas furnaces.
	The \ac{SPF} cannot be estimated without such an assumption (see \autoref{eq:E_h.j}).
	The methodology can be applied to any smart meter dataset containing separate time series of heat pumps and gas consumption in the same region.
}	

	% for what can it be used? give some examples
	The results of this paper can be used to predict the electricity consumption of future retrofitted heat pumps. \deleted{on the building and city levels.} %if gas consumption data is available.
\added{
	On the one hand, such a prediction is helpful for analyses on the building level like \cite{2023_DecarbonizationOfResidentialHeatingBasedOnSMD,2023_Wamburu_HP_replacment_with_data} or for developing city-scale digital twins like presented in \cite{2023_Bayer_DT_LocalEnergySystem}.
}
\deleted{
	On the one hand, these results are helpful whenever analyses on heat pumps retrofit on building level are carried out.
	This happens for example in the related work [10, 21] or is important for developing city-scale digital twins like presented in [2].
}
	On the other hand, utility companies can use this methodology to estimate the electricity demand produced by retrofitted heat pumps.
	This knowledge can help to optimally dimension future distribution systems and thus help to save resources on our planet as companies can avoid wrong energy system dimensioning.
	Generally, this paper shows that smart meter data of gas and electricity consumption becomes very valuable when predicting the electricity demand of retrofitted heat pumps.

\section{Conclusion and Outlook}
\label{sec:conclusion}
	%What are the main outcomes of the paper?
	This work presents a novel method to estimate the \ac{SPF} %or the \ac{SPF} divided through the mean heat demand reduction rate
	based on an unpaired dataset of energy consumption data of buildings with heat pumps and others with gas furnaces.
	The \ac{SPF} can be computed based on minimizing the \ac{JSD} of these two unpaired distributions.
	We apply our method to an exemplary, recent dataset and validate the results with related publications.
	Moreover, we estimate the impact heat pumps would have on the local distribution system and on an individual level if they were to replace existing gas furnaces.
	
	% Why is this paper useful?
	As discussed in the introduction, heat pumps will continuously replace gas furnaces in the future.
	Researchers can use the presented results to predict the impact of a higher share of retrofitted heat pumps replacing existing gas furnaces.
	The knowledge of the \ac{SPF}, which is highly dependent on climatic conditions, is fundamental for predicting the electricity demand of a heat pump that should replace an existing gas furnace on the individual building level.
	Using average values will lead to high deviations, as our analysis demonstrated for different years within the same dataset.
	
	% open questions / future work
	An open question is how retrofitted heat pumps reduce the carbon dioxide footprint when considering the complete life cycle compared to gas furnaces, including the retrofit emissions.
	Based on our approach, the sustainability of heat pump retrofit can be assessed in a realistic setting.
	% discussion on SPF depending on the control
	Another more technical, open question is how the average \ac{SPF} will change with the ongoing development of smart grids.
	More sophisticated control strategies for heat pumps that try, e.g., to maximize \ac{PV} self-consumption like \cite{2021_Yousefi_MPC_for_HP_EV_selfcons_opti} or try to minimize the overall heat demand like \cite{2022_Bayer_Enhancing_RL_for_HVAC} might change the \ac{SPF} of an individual heat pump, as highlighted by \cite{2019_Pospisil_COP_Analysis_MPC_Control}.
	%In \cite{2019_Pospisil_COP_Analysis_MPC_Control}, the authors analyze that different control strategies like maximizing \ac{PV} self-consumption or local wind production impact the seasonal \ac{COP}.
	%Depending on the strategy, increased up to 14\% but also decreases about 1.5\% are reported.
	% this is bad!

%\section{Appendices}
%
%If your work needs an appendix, add it before the ``\verb|\end{document}|'' command at the conclusion of your source document.

%Start the appendix with the ``\verb|appendix|'' command:
%\begin{verbatim}
%  \appendix
%\end{verbatim}
%and note that in the appendix, sections are lettered, not numbered. This document has two appendices, demonstrating the section and subsection identification method.

\section*{Acknowledgments}
	We would like to thank our research partner Stadtwerk Hassfurt GmbH for providing the data set.
	This paper is an outcome of the research project \textit{DigiSWM} (DIK-2103-0017 / DIK0298/02) founded by the Bavarian State Ministry of Economic Affairs, Regional Development and Energy.

%\bibliographystyle{ACM-Reference-Format}
\bibliographystyle{acm}
\bibliography{references}

%\appendix


\end{document}